\documentclass{aip-cp}

\usepackage[numbers]{natbib}
\usepackage{rotating}
\usepackage{graphicx}
\usepackage{units}
\usepackage{xfrac}
\usepackage{tabularx}
\usepackage{siunitx}
\usepackage{multirow}
\usepackage{array}
\usepackage{placeins}
\usepackage{textcomp}

\newcommand{\specialcell}[2][c]{%
  \begin{tabular}[#1]{@{}c@{}}#2\end{tabular}}

\begin{document}

\title{Constraining Radon Backgrounds in LZ}

\author[SDSMT]{E.\,H. Miller\corref{cor1}}
\author[UA]{J. Busenitz}
\author[UMd]{T.\,K. Edberg}
\author[UCL]{C. Ghag}
\author[UMd]{C. Hall}
\author[SDSMT]{R. Leonard}
\author[LBNL]{K. Lesko}
\author[UCL]{X. Liu}
\author[UA]{Y. Meng}
\author[UA]{A. Piepke}
\author[SDSMT]{R.\,W. Schnee}

\affil[SDSMT]{Department of Physics, South Dakota School of Mines \& Technology, Rapid City, SD 57701 USA}
\affil[UA]{University of Alabama, Department of Physics \& Astronomy, 206 Gallalee Hall, 514 University Boulevard, Tuscaloosa, AL 34587-0324, USA}
\affil[UMd]{University of Maryland, Department of Physics, College Park, MD 20742-4111, USA}
\affil[UCL]{University College London (UCL), Department of Physics and Astronomy, Gower Street, London, WC1E 6BT, UK}
\affil[LBNL]{Lawrence Berkeley National Laboratory (LBNL), 1 Cyclotron Road, Berkeley, CA 94720-8099, USA}
\corresp[cor1]{Corresponding author: eric.miller@sdsmt.edu}

\maketitle

\begin{abstract}
The LZ dark matter detector, like many other rare-event searches, will suffer from backgrounds due to the radioactive decay of radon daughters.  
In order to achieve its science goals, the concentration of radon within the xenon should not exceed 2\,\si{\micro}Bq/kg, or 20 mBq total within its 10 tonnes.  
The LZ collaboration is in the midst of a program to screen all significant components in contact with the xenon.
The four institutions involved in this effort have begun sharing two cross-calibration sources to ensure consistent measurement results across multiple distinct devices.  
We present here five preliminary screening results, some mitigation strategies that will reduce the amount of radon produced by the most problematic components, and a summary of the current
estimate of radon emanation throughout the detector.  
This best estimate totals $<17.3$ mBq, sufficiently low to meet the detector's science goals.  
\end{abstract}


\section{INTRODUCTION}


Radon and radon daughters pose significant, and often dominant, backgrounds to low-background and rare-event search experiments.  
These include detectors of neutrinos \cite{BOREXINO}, neutrinoless double-beta decay \cite{EXO}, and dark matter \cite{LUX, XENON1T, DRIFT, LZTDR}.  
Within the LUX-Zeplin (LZ) detector, radon and radon daughters can diffuse into the fiducial volume within the liquid xenon.  
Here, radon's prompt granddaughter $^{214}$Pb decays by a beta emission that may be mistaken for a nuclear recoil $\sim0.5\%$ of the time \cite{LZTDR}.  
This decay is separated in time from its parent and daughter decays with half-lives of 26.8 min and 19.9 min respectively, and it occurs without any associated gamma emission 9.2\% of the time.  
These factors together yield a background that may be indistinguishable from a dark-matter-induced nuclear recoil, that is distributed evenly throughout the fiducial volume, and that cannot be tagged on an event-by-event basis.  
Events derived from radon are expected to be the single largest background in the LZ detector based on a projected activity of 1.73 \si{\micro}Bq/kg, 
accounting for 3.66 of the 6.29 total WIMP-like events over a 1,000 day run \cite{LZTDR}.

\section{MITIGATION OF RADON SOURCES IN LZ}

The LZ experiment will rely on two significant mitigation strategies that reduce the amount of radon produced by certain components.  
In addition, the low temperature of parts in contact with the liquid xenon is known to reduce the diffusion of radon through materials, which can reduce the amount of radon emanated.  
For sufficiently thick materials, the amount of radon diffusing out of a material is proportional to $\sqrt{D(T)\tau}$, 
where $\tau=5.5$ days is the lifetime of $^{222}$Rn and $D(T)$ follows the Arrhenius relation \cite{Arrhenius,AndreasKirill}: 
$D(T) \propto \exp(-T_0/T)$ 
where $T_0$ is a temperature corresponding to the characteristic energy of diffusion for radon through a particular material.  
This dependence can result in significantly lower rates of radon emanation at lower temperatures in materials dominated by diffusion, like plastics 
(however, this has little or no effect on surface contamination including dust).  
For example Fluorinated Ethylene Propylene (FEP) has $T_0 \approx 6800$\textdegree K \cite{AndreasKirill, PaulyHoecht}, 
so the diffusion coefficient $D(T)$ at 161\textdegree K is reduced by a factor of $2.5\times10^8$ from room temperature at 298\textdegree K, 
resulting in a reduction in the radon production by about four orders of magnitude.  
Radon emanation measurements (described below) performed at room temperature therefore significantly over-estimate the contribution of plastics that will be cold within the LZ detector.  
LZ will take advantage of this feature by cladding all cables in FEP, or a similar plastic, to effectively block radon produced below the cable's surface from diffusing into the xenon.  

\begin{figure}[t]
  \centerline{\includegraphics[width=0.95\textwidth]{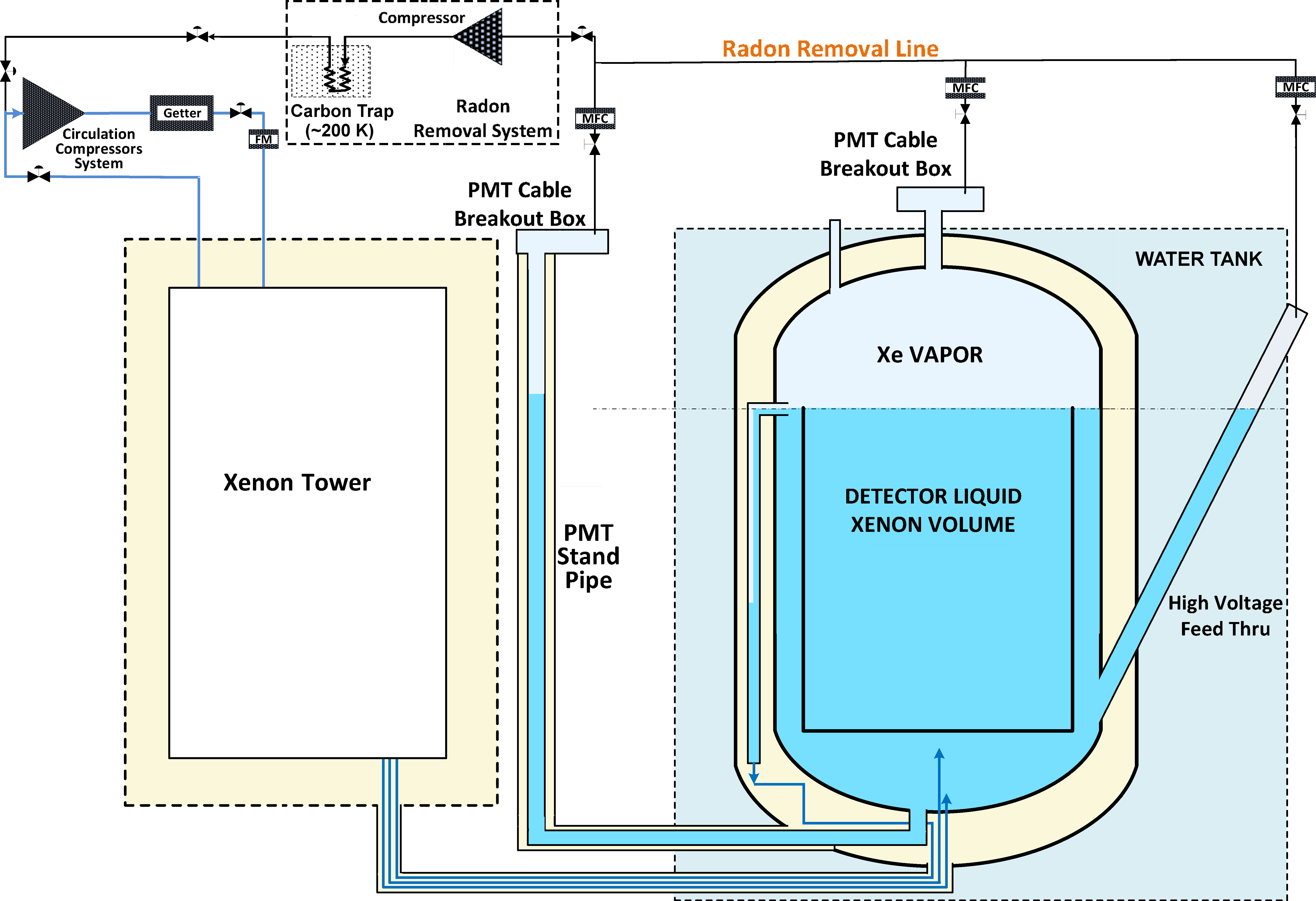}}
  \caption{A simplified schematic of the xenon recirculation in LZ showing the primary recirculation path through the xenon tower (left).  
  The circulation path from the top and bottom PMT breakout boxes and from the high voltage feedthrough passes through the radon removal system (top middle) before re-entering the main recirculation.  
  Here the vacuum space is yellow, the liquid xenon is blue, the gaseous xenon is pale blue, the main liquid recirculation is dark blue lines, and the main gaseous recirculation is light blue lines.  
  A full version can be found in \cite{LZTDR}.  }
  \label{fig:Circulation}
\end{figure}

This strategy should significantly reduce the amount of radon produced by the sections of cable that are immersed in liquid xenon and held at 161\textdegree K. 
The ends of the cables where they meet the feedthroughs to lab space, however, are at room temperature and still produce radon at a higher rate.  
In order to mitigate this source of radon, LZ will employ a cooled-carbon trap, similar to \cite{CarbonTrap}.  
While it is impractical to use such a trap to filter out the full recirculation of xenon (planned at 500 slpm), 
LZ is developing a device capable of removing 90\% of the radon from a modest flow of only 1 slpm xenon \cite{LZTDR}.  
This system will purify the xenon gas from low-flow regions of the detector before letting it re-enter the main xenon flow.  
As shown in Figure \ref{fig:Circulation}, the regions affected by this system include the gaseous spaces around the umbilical cable as well as cables from both the top and bottom PMTs, including all cable feedthroughs.  

The feedthroughs carrying HV for the detector's PMTs contact the xenon in a space purified by the cooled carbon trap, so only 10\% of the radon produced will affect LZ.
However, the insulator on these feedthroughs is an alumina ceramic with the potential to be high in uranium, and therefore high in radon production.  
These feedthroughs, and potentially other components, will be encapsulated in 3 mm of MasterBond epoxy.
This epoxy, demonstrated by BOREXINO to be low-radioactivity \cite{Borexino-Epoxy}, will create a barrier separating the feedthrough ceramic from the xenon gas and significantly reducing its radon contribution.  
The effectiveness of this epoxy to prevent radon emanation is currently under measurement.  

\section{RADON EMANATION MEASUREMENTS}
Four institutions within the LZ collaboration have developed systems to measure the amount of radon produced by items:
University of Alabama (UA) employs a system that dissolves Rn into liquid scintillator and identifies radon by the $^{214}$Bi-$^{214}$Po timing coincidence; 
University College London (UCL) \cite{SuperNEMORn}, University of Maryland (UMd), and South Dakota School of Mines and Technology (SDSM\&T)
use systems where charged radon daughters are collected on a silicon-pin diode which measures the subsequent alpha decays.  
These are summarized in Table \ref{tab:Systems}.  

With four institutions making measurements, using two different measurement techniques, ensuring consistent measurements is critical.  
Each system has been individually calibrated by adding known amounts of radon to a vessel via a calibrated flow-through source.  
In addition, LZ is engaging in a cross-calibration campaign between the various institutions in which two different samples are being measured in each emanation chamber.
The first sample is a strip of rubber that produces $\mathcal{O}(10)$ mBq (generously lent by J. Farine from work with EXO \cite{EXO}), testing the systems' calibrated measurement efficiency. 
The second sample, a set of thoriated TIG welding rods, produces $\mathcal{O}(1)$ mBq and tests the systems' background measurements --- calibration with this sample will begin soon.

Each institution's measurements, relative to EXO's original measurements, are shown in Table \ref{tab:Systems}.  
Most institutions agree to within uncertainties; UCL's anomalous result is under further study.  
The emanation and measurement process for the SDSM\&T system is described briefly below.  

\begin{table}[ht]
\centering
 \begin{tabular}{|c|c|c|c|c|c|}
\noalign{\hrule height 0.5pt}
 Institution & Technology & \specialcell{Sample \\ Volume} & Blank Rate  & \specialcell{Measurement \\(relative to EXO)} & \specialcell{LZ Sample \\ Throughput}\\
\noalign{\hrule height 1.0pt}
 UCL & \specialcell{Electrostatic \\ PIN-diode} & \specialcell{2.6 liters  \\ 2.6 liters} & \specialcell{0.2 mBq \\ 0.4 mBq} & \specialcell{$1.49 \pm 0.05$ \\ \, } & 6 / year \\
\noalign{\hrule height 0.5pt}
 UMd & \specialcell{Electrostatic \\ PIN-diode} & 4.7 liters & 0.2 mBq & $1.13 \pm 0.06 \pm 0.18$ & 12 / year \\
\noalign{\hrule height 0.5pt}
 SDSM\&T & \specialcell{Electrostatic \\ PIN-diode} & \specialcell{13 liters \\ 300 liters} & \specialcell{$<$0.3 mBq \\ 0.3 mBq} & \specialcell{$0.89\pm0.12\pm0.15$ \\ \,}& 18 / year \\
\noalign{\hrule height 0.5pt}
 UA & \specialcell{Liquid Scintillator \\ Coincidence} & \specialcell{ 2.6 liters  \\ 2.6 liters} & \specialcell{0.2 mBq \\ 0.2 mBq} & \specialcell{$0.83 \pm 0.15 \pm 0.08$ \\ $0.85 \pm 0.19 \pm 0.08$} & 24 / year \\
\noalign{\hrule height 0.5pt}
 \end{tabular}
\caption{Seven vessels, at four institutions, are available to screen LZ samples for radon emanation with the throughputs shown. 
Most of the measurements are consistent within quoted errors.
  SDSM\&T's 300\,L and UCL's second 2.6\,L vessels are still awaiting calibration.  
  }
\label{tab:Systems}
\end{table}

Sample preparation includes a hand-scrub of the entire surface with isopropyl alcohol-soaked lint-free wipes in order to remove surface dust.  
The sample is then sealed in a vacuum-tight emanation chamber, flushed with boil-off nitrogen, evacuated to near-vacuum, and allowed to out-gas before a measurement. 
This outgasing time ranges from a few days for metallic samples to weeks for porous samples --- 
these typically absorb atmospheric radon which can subsequently diffuse out of the sample during a measurement, overwhelming the contribution intrinsic to the sample itself.  
Finally, the chamber is flushed with boil-off nitrogen and filled to $\approx100$ Torr.  
An activated charcoal trap, cooled to -77\textdegree C by a bath of isopropyl alcohol and dry ice, further reduces the radon content of the boil-off nitrogen.
Use of 100 Torr rather than vacuum ensures that radon atoms produced by the nuclear recoil of a $^{226}$Ra decay are captured in the gas, 
rather than embedding in the metallic wall of the emanation chamber.  
An 86.3 keV $^{222}$Rn atom, produced by the decay of $^{226}$Ra, 
has a stopping range of 5 meters at a modest vacuum of 0.1 Torr compared with 500 \si{\micro\m} in 100 Torr nitrogen gas \cite{SRIM}.



\begin{figure}[ht]
  \begin{tabularx}{\textwidth}{l r}
    \includegraphics[width=0.5\textwidth]{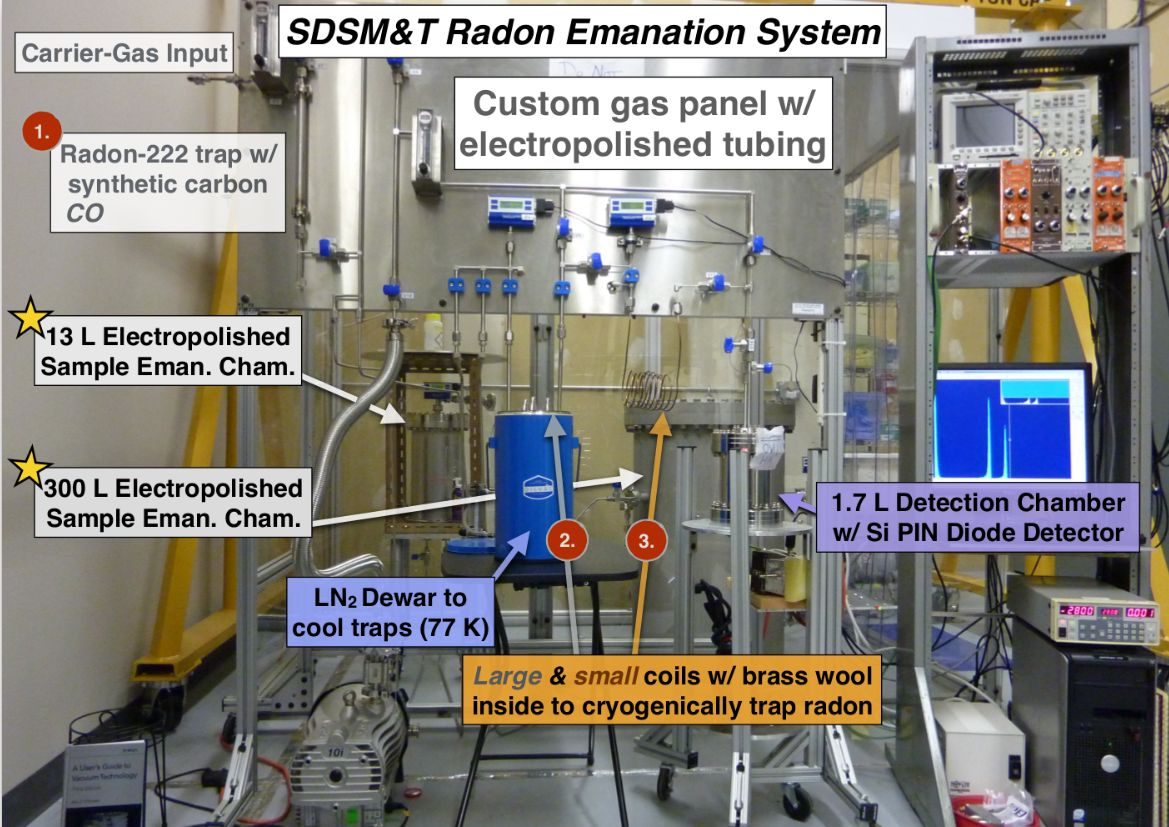} &
    \includegraphics[width=0.45\textwidth]{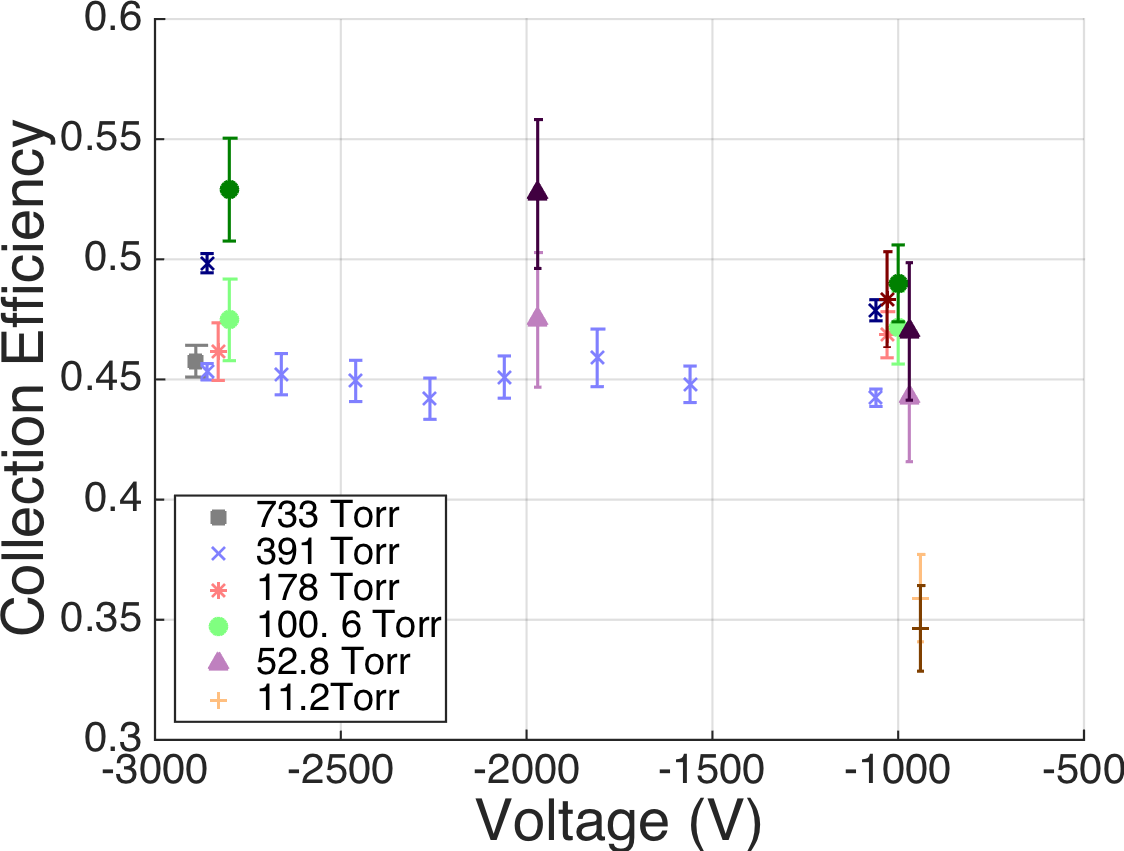}
    \end{tabularx}
    \caption{
    \textit{Left}: The radon emanation measurement system at SDSM\&T -- components and operation described in text.
    \textit{Right}: The efficiency for collecting radon daughters onto the silicon-pin diode in a nitrogen atmosphere does not vary significantly with changes in voltage or pressure.
    Lighter points correspond to collection efficiency for $^{218}$Po, while the darker points are for $^{214}$Po.  
    SDSM\&T operates at around 100 Torr and -2000 V.  
    }
    \label{fig:twofigs}
\end{figure}

As shown in Figure \ref{fig:twofigs} left, SDSM\&T employs two electropolished stainless steel vessels for radon emanation --- one 13\,L, and one 300\,L.  
Each sample is allowed to emanate in a vessel for around one week.  
During this time, the activity of radon within the chamber ($C(t)$) begins to approach the equilibrium activity produced by the material 
($C_0$) as $C(t) = C_0\left(1-\exp\left(\sfrac{-t}{\tau}\right)\right)$, where $\tau = 5.5$ days is the $^{222}$Rn lifetime.

After this emanation period the nitrogen gas, and the radon contained therein, is evacuated from the emanation chamber through a cold trap.  
During this process, the pressure inside the emanation chamber drops to $\approx20$ Torr, and is refilled back to $\approx100$ Torr, four times.  
This pressure cycling helps to ensure that the vast majority of radon from the chamber is captured, and that there are no pockets of gas that 
might 
remain isolated.  
The cold trap is a length of \sfrac{3}{8}'' electropolished stainless-steel tubing filled with brass wool; this tubing is submersed in a liquid nitrogen bath so that
radon atoms adsorb to the surface of the brass wool.  
Since the pressure within this trap remains low throughout the process, the nitrogen carrier gas does not condense within the trap.  

Once the radon has been frozen into this first, large trap, the trap is warmed back to atmospheric pressure and a flow of fresh boil-off nitrogen gas carries the radon from this large trap
into a second, smaller cold trap whose volume is small compared to that of the detection chamber.  
This second trap is made from \sfrac{1}{8}'' copper tubing and is also filled with brass wool and cooled with liquid nitrogen.  
Finally, the small copper trap is warmed to room temperature, and the 1.3 liter detection chamber is filled to 100 Torr with fresh boil-off nitrogen flowing through the trap.
This process robustly and repeatably transfers at least 95\% of the radon from the emanation chamber into the detection chamber.  

The detection chamber is a 1.3 liter electropolished stainless-steel vessel containing a 1'' silicon-pin diode held at -2000 V.  
Following the decay of $^{222}$Rn atoms, $\approx88\%$ of the $^{218}$Po daughters are positively charged and follow the electric field lines to the diode \cite{ChargedFraction}.  
As shown in Figure \ref{fig:twofigs} right, the probability for collecting for each prompt radon daughter onto the diode is $\approx45$\%.  
The subsequent alpha decays of $^{218}$Po and $^{214}$Po produce clear signals in the diode, which is read out by a multi-channel analyzer.

\section{PRELIMINARY RESULTS}

The radon emanation screening program for LZ is now well underway with 63 measurements completed.  
Table \ref{tab:Results} lists five preliminary measurements, along with calculations of the components' total contributions within the LZ experiment
with and without the mitigation strategies discussed above.  
The 17 km of Axon cable, used to transfer high voltage to and signals from the detector's PMTs, would produce 15.7 mBq at room temperature.  
This contribution alone would approach the allowable limit within the detector.
Accounting for the reduction by both the low temperature of the sections within the liquid xenon and the radon removal by the cooled carbon trap results in 
the modest estimate of 0.8 mBq.  



\begin{table}[ht]
\centering
 \begin{tabular}{|c|c|c|c|c|c|}
\noalign{\hrule height 0.5pt}
  Material & Vessel & Result & Units & Full Contribution & Mitigated Contribution \\
\noalign{\hrule height 1.0pt}
  Axon Cable & SDSM\&T 300\,L & $0.93\pm0.27$ & mBq/km & $15.7 \pm 4.5$ mBq & $0.8 \pm 0.23$ mBq \\
\noalign{\hrule height 0.5pt}
  HV Feedthrus & UA 2.6\,L & $0.5\pm0.2$ & mBq/unit & $6\pm2.4$ mBq & $0.6\pm0.24$ mBq \\
\noalign{\hrule height 0.5pt}
  PMT Bases & UCL 2.6\,L & $0.28\pm0.17$ & mBq/unit & $1.8\pm1.1$ mBq & $1.8\pm1.1$ mBq$^\ast$ \\
\noalign{\hrule height 0.5pt}
  PTFE & SDSM\&T 300\,L & $<0.015$ & mBq/m$^2$ & $<1.29$ mBq & $<1.29$ mBq$^\ast$ \\
\noalign{\hrule height 0.5pt}
  Umbilical Cable & SDSM\&T 13\,L & $0.26\pm0.06$ & mBq/m & $2.1\pm0.5$ mBq & Rejected \\
\noalign{\hrule height 0.5pt}
  
  \end{tabular}
  \caption{Preliminary results from the LZ radon emanation screening program.  
    The full contribution is the amount of radon produced by the total amount of each material in LZ, 
    while the mitigated contribution accounts for radon reduction by the cooled carbon trap and, in the case of the Axon cables, a reduction due to temperature. 
    Items indicated with $^\ast$ are expected to produce less radon by an unknown amount due to low temperature.  
    This umbilical cable listed was rejected and will not be used.}
  \label{tab:Results}
\end{table}

The 20 mBq requirement combined with the large total surface area exposed to xenon within LZ 
requires the measurements of samples with large surface area.  
One example is PTFE which, making up the TPC and reflector lining the inner cryostat, will have an area 84 m$^2$.  
The first screening sample of this material was 18 m$^2$ and fit only within the collaboration's largest emanation chamber, the 300L vessel at SDSM\&T.  
The result of screening even this large a sample was consistent with zero, with a 90\% upper limit of 0.015 mBq/m$^2$.

An additional significant contributor of radon within LZ will be dust on surfaces.  
With only 500 ng/cm$^2$ on xenon-wetted surfaces, the experiment will hold about 1 gram of dust.  
Assuming typical activity of $\approx 40$ mBq/g of uranium, and a conservative estimate that 25\% of the radon produced by surface dust escapes into xenon, 
this contaminant will contribute 10 mBq of radon to the experiment.

\begin{table}[t]
  \centering
  \begin{tabular}{|c|c|r|c|c|}
    \noalign{\hrule height 0.5pt}
    \textbf{Material} & \textbf{Component(s)} & \textbf{Quantity} & \textbf{Unit}  & \begin{tabular}[c]{@{}c@{}} \textbf{Estimate} \\ \textbf{(mBq)}\end{tabular} \\ 
    \noalign{\hrule height 1.0pt}
    Al$_2$O$_3$ resistor & PMT Bases & 9790 & \# & \textbf{0.58}$^\ast$ \\ 
    BaTiO$_3$ capacitor & PMT Bases & 3010 & \# &  \textbf{0.016}$^\ast$ \\ 
    Cirlex & PMT Bases & 6000 & \si{\cm\squared} &  \textbf{0.37}$^\ast$\\ 
    \noalign{\hrule height 0.5pt}
    Titanium & \specialcell{Cryostat, PMT Mounts,\\ Field Rings, Grid Supports}  & 412,000 & \si{\cm\squared} & 0.41\\ 
    \noalign{\hrule height 0.5pt}
    PTFE &Reflectors, Field Cage & 840,000 & \si{\cm\squared}  & $<$\textbf{1.3}$^\ast$\\ 
    \noalign{\hrule height 0.5pt}
    PMT cabling$^\dagger$ & PMT Cabling & 17,000 & \si{\m} &  \textbf{0.8} \\ 
    PMT feedthrough$^\dagger$ & PMT Feedthrough & 88 & \# &  $<$\textbf{0.24} \\ 
    Steel conduit$^\dagger$ & Cabling Conduit & 100,000 & \si{\cm\squared} & 0.055\\ 
    R11410 PMT & R11410 PMT & 488 & \# &  1.26 \\ 
    R8520 PMT & R8520 PMT & 90 & \# &  0.15 \\ 
    R8778 PMT & R8778 PMT & 38 & \# & 0.09 \\
    Polyethylene & HV Umbilical & 4200 & \si{\cm\squared} &  0.10$^\ast$\\ 
    Tin-coated copper & HV Umbilical & 11,000 & \si{\cm\squared}  & 0.002 \\ 
    Tivar & HV Umbilical & 3894 & \si{\cm\squared} &  0.004$^\ast$\\ 
    Acetal & HV Umbilical & 195 & \si{\cm\squared} &  0.0002$^\ast$\\ 
    Copper & HV Umbilical & 39 & \si{\cm\squared} & 0.000007 \\ 
    Epoxy & HV Umbilical, Feedthroughs & 1000 & \si{\cm\squared}  & 0.0001$^\ast$\\ 
    \noalign{\hrule height 0.5pt}
    Steel & Cryostat Seals, Xe Recirculation  & 135,000 & \si{\cm\squared} &  0.104\\ 
    \noalign{\hrule height 1.0pt}
    Recirculation pump & Xe Recirculation & 1 & \# &  0.1\\ 
    Purification getter & Xe Recirculation & 2.5 & \si{\kg} &  1.34 \\ 
    Transducers \& Valves & Xe Recirculation & 30 & \# &  0.17 \\ 
    \noalign{\hrule height 0.5pt}
    Welds & Recirculation System, Cryostat & 32.3 & m &  \textbf{0.11} \\ 
    \noalign{\hrule height 0.5pt}
    Dust &  & 1 & g &  10.0 \\ 
    \noalign{\hrule height 1.0pt}
    \textbf{Total} &  &  &  & \textbf{$<17.3$} \\ 
    \noalign{\hrule height 0.5pt}
  \end{tabular}

  \caption{List of materials in contact with Xe, indicating the quantity of the material and the goal for radon emanation from the material.
  The estimates of radon emanation are expected usually to be conservative, as they are based on the most similar object or material for which emanation rates are available in the literature, 
  and use only conservative models for reduction of radon emanation at LXe temperatures. 
  Some materials are expected to emanate less radon when cold, so room-temperature emanation quantities (labeled with $^\ast$) may be reduced. 
  Expected reduction of radon by the carbon trap is included in estimates for those components affected (labeled with $^\dagger$).  
  Values in \textbf{bold} are based on measurements performed by LZ.}
  \label{tab:Budget}
\end{table}

In order to reach its science goals, LZ must limit the activity of $^{222}$Rn within the 10 tonnes of xenon to 2 \si{\micro}Bq/kg, or 20 mBq total.  
The experiment has kept an inventory of all components that will be in contact with the xenon along with an estimate of radon emanation from each 
\cite{Zuzel2005, SNORn, Xenon2011, Zuzel2009}. 
This estimation includes measurements from the screening program described above and is shown in Table \ref{tab:Budget}.  
This radon budget helps to direct screening and mitigation efforts and informs experiment background models.  
The best estimates at time of writing indicate a contribution of 10 mBq from dust, and $<7.3$ mBq from materials for a total of $<17.3$ mBq total within the detector's xenon.  

\FloatBarrier


\section{ACKNOWLEDGMENTS}
This work was supported in part by the Department of Energy (Grants No. DE-SC0014223, DE-AC02-05CH11231, DE-SC0010072, DE-SC0012447), 
and by the the U.K. Science \& Technology Facilities Council under award number ST/M003981/1.

\FloatBarrier
\bibliographystyle{aipnum-cp}%
\bibliography{LRTBib}%

\begin{thebibliography}{18}%
\makeatletter
\providecommand \@ifxundefined [1]{%
 \@ifx{#1\undefined}
}%
\providecommand \@ifnum [1]{%
 \ifnum #1\expandafter \@firstoftwo
 \else \expandafter \@secondoftwo
 \fi
}%
\providecommand \@ifx [1]{%
 \ifx #1\expandafter \@firstoftwo
 \else \expandafter \@secondoftwo
 \fi
}%
\providecommand \natexlab [1]{#1}%
\providecommand \enquote  [1]{``#1''}%
\providecommand \bibnamefont  [1]{#1}%
\providecommand \bibfnamefont [1]{#1}%
\providecommand \citenamefont [1]{#1}%
\providecommand \href@noop [0]{\@secondoftwo}%
\providecommand \href [0]{\begingroup \@sanitize@url \@href}%
\providecommand \@href[1]{\@@startlink{#1}\@@href}%
\providecommand \@@href[1]{\endgroup#1\@@endlink}%
\providecommand \@sanitize@url [0]{\catcode `\$12\catcode `\&12\catcode
  `\#12\catcode `\^12\catcode `\_12\catcode `\%12\relax}%
\providecommand \@@startlink[1]{}%
\providecommand \@@endlink[0]{}%
\providecommand \url  [0]{\begingroup\@sanitize@url \@url }%
\providecommand \@url [1]{\endgroup\@href {#1}{\urlprefix }}%
\providecommand \urlprefix  [0]{URL }%
\providecommand \Eprint [0]{\href }%
\providecommand \doibase [0]{http://dx.doi.org/}%
\providecommand \selectlanguage [0]{\@gobble}%
\providecommand \bibinfo  [0]{\@secondoftwo}%
\providecommand \bibfield  [0]{\@secondoftwo}%
\providecommand \translation [1]{[#1]}%
\providecommand \BibitemOpen [0]{}%
\providecommand \bibitemStop [0]{}%
\providecommand \bibitemNoStop [0]{.\EOS\space}%
\providecommand \EOS [0]{\spacefactor3000\relax}%
\providecommand \BibitemShut  [1]{\csname bibitem#1\endcsname}%
\let\auto@bib@innerbib\@empty
\bibitem [{\citenamefont {Bellini}\ \emph {et~al.}(2013)\citenamefont {Bellini}
  \emph {et~al.}}]{BOREXINO}%
  \BibitemOpen
  \bibfield  {author} {\bibinfo {author} {\bibfnamefont {G.}~\bibnamefont
  {Bellini}} \emph {et~al.} (\bibinfo {collaboration} {Borexino}),\ }\href
  {\doibase 10.1088/1475-7516/2013/08/049} {\bibfield  {journal} {\bibinfo
  {journal} {JCAP}\ }\textbf {\bibinfo {volume} {1308}},\ p.\ \bibinfo {pages}
  {049} (\bibinfo {year} {2013})},\ \Eprint {http://arxiv.org/abs/1304.7381}
  {arXiv:1304.7381 [physics.ins-det]} \BibitemShut {NoStop}%
\bibitem [{\citenamefont {Albert}\ \emph {et~al.}(2015)\citenamefont {Albert}
  \emph {et~al.}}]{EXO}%
  \BibitemOpen
  \bibfield  {author} {\bibinfo {author} {\bibfnamefont {J.~B.}\ \bibnamefont
  {Albert}} \emph {et~al.} (\bibinfo {collaboration} {EXO-200 Collaboration}),\
  }\href {\doibase 10.1103/PhysRevC.92.015503} {\bibfield  {journal} {\bibinfo
  {journal} {Phys. Rev. C}\ }\textbf {\bibinfo {volume} {92}},\ p.\ \bibinfo
  {pages} {015503}Jul (\bibinfo {year} {2015})}\BibitemShut {NoStop}%
\bibitem [{\citenamefont {Bradley}\ \emph {et~al.}(2015)\citenamefont {Bradley}
  \emph {et~al.}}]{LUX}%
  \BibitemOpen
  \bibfield  {author} {\bibinfo {author} {\bibfnamefont {A.}~\bibnamefont
  {Bradley}} \emph {et~al.},\ }\href {\doibase
  http://dx.doi.org/10.1016/j.phpro.2014.12.067} {\bibfield  {journal}
  {\bibinfo  {journal} {Physics Procedia}\ }\textbf {\bibinfo {volume} {61}},\
  \unskip\ \bibinfo {pages} {658 -- 665} (\bibinfo {year} {2015})},\ \bibinfo
  {note} {13th International Conference on Topics in Astroparticle and
  Underground Physics, TAUP 2013}\BibitemShut {NoStop}%
\bibitem [{\citenamefont {Aprile}\ \emph {et~al.}(2016)\citenamefont {Aprile}
  \emph {et~al.}}]{XENON1T}%
  \BibitemOpen
  \bibfield  {author} {\bibinfo {author} {\bibfnamefont {E.}~\bibnamefont
  {Aprile}} \emph {et~al.} (\bibinfo {collaboration} {XENON}),\ }\href
  {\doibase 10.1088/1475-7516/2016/04/027} {\bibfield  {journal} {\bibinfo
  {journal} {JCAP}\ }\textbf {\bibinfo {volume} {1604}},\ p.\ \bibinfo {pages}
  {027} (\bibinfo {year} {2016})},\ \Eprint {http://arxiv.org/abs/1512.07501}
  {arXiv:1512.07501 [physics.ins-det]} \BibitemShut {NoStop}%
\bibitem [{\citenamefont {Brack}\ \emph {et~al.}(2014)\citenamefont {Brack}
  \emph {et~al.}}]{DRIFT}%
  \BibitemOpen
  \bibfield  {author} {\bibinfo {author} {\bibfnamefont {J.}~\bibnamefont
  {Brack}} \emph {et~al.} (\bibinfo {collaboration} {DRIFT}),\ }\href
  {http://stacks.iop.org/1748-0221/9/i=07/a=P07021} {\bibfield  {journal}
  {\bibinfo  {journal} {Journal of Instrumentation}\ }\textbf {\bibinfo
  {volume} {9}},\ p.\ \bibinfo {pages} {P07021} (\bibinfo {year}
  {2014})}\BibitemShut {NoStop}%
\bibitem [{\citenamefont {Mount}\ \emph {et~al.}(2017)\citenamefont {Mount}
  \emph {et~al.}}]{LZTDR}%
  \BibitemOpen
  \bibfield  {author} {\bibinfo {author} {\bibfnamefont {B.~J.}\ \bibnamefont
  {Mount}} \emph {et~al.},\ }\href@noop {} {\  (\bibinfo {year} {2017})},\
  \Eprint {http://arxiv.org/abs/1703.09144} {arXiv:1703.09144
  [physics.ins-det]} \BibitemShut {NoStop}%
\bibitem [{\citenamefont {Arrhenius}(1889)}]{Arrhenius}%
  \BibitemOpen
  \bibfield  {author} {\bibinfo {author} {\bibfnamefont {S.}~\bibnamefont
  {Arrhenius}},\ }\href@noop {} {\emph {\bibinfo {title} {{\"U}ber die
  Dissociationsw{\"a}rme und den Einfluss der Temperatur auf den
  Dissociationsgrad der Elektrolyte}}}\ (\bibinfo  {publisher} {Wilhelm
  Engelmann},\ \bibinfo {address} {Leipzig},\ \bibinfo {year} {1889})\ \unskip,
  pp.\ \bibinfo {pages} {96--116}\BibitemShut {NoStop}%
\bibitem [{\citenamefont {Piepke}\ and\ \citenamefont
  {Pushkin}(2011)}]{AndreasKirill}%
  \BibitemOpen
  \bibfield  {author} {\bibinfo {author} {\bibfnamefont {A.}~\bibnamefont
  {Piepke}}\ and\ \bibinfo {author} {\bibfnamefont {K.}~\bibnamefont
  {Pushkin}},\ }\href@noop {} {\enquote {\bibinfo {title} {Estimate of the
  {EXO}-200 radon production inside the {TPC}},}\ } (\bibinfo {year} {2011}),\
  \bibinfo {note} {{EXO} Internal Memo}\BibitemShut {NoStop}%
\bibitem [{\citenamefont {Pauly}()}]{PaulyHoecht}%
  \BibitemOpen
  \bibfield  {author} {\bibinfo {author} {\bibfnamefont {S.}~\bibnamefont
  {Pauly}},\ }\href@noop {} {\enquote {\bibinfo {title} {Permeability and
  diffusion data},}\ }\bibinfo {note} {Hoechst AG, Werke Kalle}\BibitemShut
  {NoStop}%
\bibitem [{\citenamefont {Abe}\ \emph {et~al.}(2012)\citenamefont {Abe} \emph
  {et~al.}}]{CarbonTrap}%
  \BibitemOpen
  \bibfield  {author} {\bibinfo {author} {\bibfnamefont {K.}~\bibnamefont
  {Abe}} \emph {et~al.},\ }\href {\doibase
  http://dx.doi.org/10.1016/j.nima.2011.09.051} {\bibfield  {journal} {\bibinfo
   {journal} {Nuclear Instruments and Methods in Physics Research Section A:
  Accelerators, Spectrometers, Detectors and Associated Equipment}\ }\textbf
  {\bibinfo {volume} {661}},\ \unskip\ \bibinfo {pages} {50 -- 57} (\bibinfo
  {year} {2012})}\BibitemShut {NoStop}%
\bibitem [{\citenamefont {Arpesella}\ \emph {et~al.}(2002)\citenamefont
  {Arpesella} \emph {et~al.}}]{Borexino-Epoxy}%
  \BibitemOpen
  \bibfield  {author} {\bibinfo {author} {\bibfnamefont {C.}~\bibnamefont
  {Arpesella}} \emph {et~al.} (\bibinfo {collaboration} {BOREXINO}),\ }\href
  {\doibase 10.1016/S0927-6505(01)00179-7} {\bibfield  {journal} {\bibinfo
  {journal} {Astropart. Phys.}\ }\textbf {\bibinfo {volume} {18}},\ \unskip\
  \bibinfo {pages} {1--25} (\bibinfo {year} {2002})},\ \Eprint
  {http://arxiv.org/abs/hep-ex/0109031} {arXiv:hep-ex/0109031 [hep-ex]}
  \BibitemShut {NoStop}%
\bibitem [{\citenamefont {Soule}(2013)}]{SuperNEMORn}%
  \BibitemOpen
  \bibfield  {author} {\bibinfo {author} {\bibfnamefont {B.}~\bibnamefont
  {Soule}},\ }\href@noop {} {\enquote {\bibinfo {title} {Radon emanation
  chamber : High sensitivity measurements for the supernemo experiment},}\ }\
  08 (\bibinfo {year} {2013})\unskip\BibitemShut {NoStop}%
\bibitem [{\citenamefont {Zeigler}(2008)}]{SRIM}%
  \BibitemOpen
  \bibfield  {author} {\bibinfo {author} {\bibfnamefont {J.}~\bibnamefont
  {Zeigler}},\ }\href@noop {} {\bibinfo {title} {The stopping ranges of ions in
  matter},\ }\  \bibinfo {year} {2008} \unskip\BibitemShut {NoStop}%
\bibitem [{\citenamefont {Porstend{\"o}rfer}(2001)}]{ChargedFraction}%
  \BibitemOpen
  \bibfield  {author} {\bibinfo {author} {\bibfnamefont {J.}~\bibnamefont
  {Porstend{\"o}rfer}},\ }\href {\doibase 10.1093/oxfordjournals.rpd.a006512}
  {\bibfield  {journal} {\bibinfo  {journal} {Radiation Protection Dosimetry}\
  }\textbf {\bibinfo {volume} {94}},\ \unskip\ \bibinfo {pages} {365--373}
  (\bibinfo {year} {2001})}\BibitemShut {NoStop}%
\bibitem [{\citenamefont {{Zuzel}}(2005)}]{Zuzel2005}%
  \BibitemOpen
  \bibfield  {author} {\bibinfo {author} {\bibfnamefont {G.}~\bibnamefont
  {{Zuzel}}},\ }\enquote {\bibinfo {title} {{Highly Sensitive Measurements of
  $^{222}$Rn Diffusion and Emanation}},}\ in\ \href {\doibase
  10.1063/1.2060465} {\emph {\bibinfo {booktitle} {Topical Workshop on Low
  Radioactivity Techniques: LRT 2004.}}},\ \bibinfo {series} {American
  Institute of Physics Conference Series}, Vol.\ \bibinfo {volume} {785},\
  \bibinfo {editor} {edited by\ \bibinfo {editor} {\bibfnamefont
  {B.}~\bibnamefont {{Cleveland}}}, \bibinfo {editor} {\bibfnamefont
  {R.}~\bibnamefont {{Ford}}}, \ and\ \bibinfo {editor} {\bibfnamefont
  {M.}~\bibnamefont {{Chen}}}}\ (\bibinfo {year} {2005})\ \unskip, pp.\
  \bibinfo {pages} {142--149}\BibitemShut {NoStop}%
\bibitem [{\citenamefont {Liu}, \citenamefont {Lee},\ and\ \citenamefont
  {McDonald}(1993)}]{SNORn}%
  \BibitemOpen
  \bibfield  {author} {\bibinfo {author} {\bibfnamefont {M.}~\bibnamefont
  {Liu}}, \bibinfo {author} {\bibfnamefont {H.}~\bibnamefont {Lee}}, \ and\
  \bibinfo {author} {\bibfnamefont {A.}~\bibnamefont {McDonald}},\ }\href
  {\doibase http://dx.doi.org/10.1016/0168-9002(93)90948-H} {\bibfield
  {journal} {\bibinfo  {journal} {Nuclear Instruments and Methods in Physics
  Research Section A: Accelerators, Spectrometers, Detectors and Associated
  Equipment}\ }\textbf {\bibinfo {volume} {329}},\ \unskip\ \bibinfo {pages}
  {291 -- 298} (\bibinfo {year} {1993})}\BibitemShut {NoStop}%
\bibitem [{\citenamefont {Aprile}\ \emph {et~al.}(2011)\citenamefont {Aprile}
  \emph {et~al.}}]{Xenon2011}%
  \BibitemOpen
  \bibfield  {author} {\bibinfo {author} {\bibfnamefont {E.}~\bibnamefont
  {Aprile}} \emph {et~al.} (\bibinfo {collaboration} {XENON100
  Collaboration}),\ }\href {\doibase 10.1103/PhysRevD.83.082001} {\bibfield
  {journal} {\bibinfo  {journal} {Phys. Rev. D}\ }\textbf {\bibinfo {volume}
  {83}},\ p.\ \bibinfo {pages} {082001}Apr (\bibinfo {year}
  {2011})}\BibitemShut {NoStop}%
\bibitem [{\citenamefont {Zuzel}\ and\ \citenamefont
  {Simgen}(2009)}]{Zuzel2009}%
  \BibitemOpen
  \bibfield  {author} {\bibinfo {author} {\bibfnamefont {G.}~\bibnamefont
  {Zuzel}}\ and\ \bibinfo {author} {\bibfnamefont {H.}~\bibnamefont {Simgen}},\
  }\href {\doibase http://dx.doi.org/10.1016/j.apradiso.2009.01.052} {\bibfield
   {journal} {\bibinfo  {journal} {Applied Radiation and Isotopes}\ }\textbf
  {\bibinfo {volume} {67}},\ \unskip\ \bibinfo {pages} {889 -- 893} (\bibinfo
  {year} {2009})},\ \bibinfo {note} {5th International Conference on
  Radionuclide Metrology - Low-Level Radioactivity Measurement Techniques
  ICRM-LLRMT'08}\BibitemShut {NoStop}%
\end{thebibliography}%

\end{document}